\documentclass{PoS}

\usepackage{amsfonts,amssymb,amsmath,amsthm,amscd,latexsym}

\newcommand{\su}{\mathfrak{su}}
\newcommand{\s}{\mathfrak{sl}}
\newcommand{\SU}{\mathrm{SU}}

\newcommand{\SL}{\mathrm{SL}}

\newcommand{\Hom}{\mathrm{Hom}}
\newcommand{\End}{\mathrm{End} \,}
\newcommand{\C}{\mathbb{C}}
\newcommand{\R}{\mathbb{R}}

\newcommand{\beq}{\begin{equation}}
\newcommand{\eeq}{\end{equation}}
\newcommand{\beqa}{\begin{eqnarray}}
\newcommand{\eeqa}{\end{eqnarray}}
\newcommand{\nn}{\nonumber}

\newtheorem{theorem}{Theorem}[section]
\newtheorem{lemma}[theorem]{Lemma}

\newtheorem{proposition}[theorem]{Proposition}

\newtheorem{definition}[theorem]{Definition}

\title{q-Deformation of Lorentzian spin foam models}

\ShortTitle{Quantum deformation of Lorentzian spin foam models}

\author{\speaker{Winston J. Fairbairn} \\    
Department of Mathematics \\ FAU Erlangen-N\"urnberg \\ Cauerstrasse 11 \\ 91054 Erlangen \\ Germany \\
        E-mail: \email{winston.fairbairn@math.uni-erlangen.de}}

\author{Catherine Meusburger\\
Department of Mathematics \\ FAU Erlangen-N\"urnberg \\ Cauerstrasse 11 \\ 91054 Erlangen \\ Germany \\
       E-mail: \email{catherine.meusburger@math.uni-erlangen.de}}

\abstract{We construct and analyse a quantum deformation of the Lorentzian EPRL model. The model is based on the 
representation theory of the quantum Lorentz group with real deformation parameter. We give a definition of the quantum 
EPRL intertwiner, study its convergence and braiding properties and construct an amplitude for the four-simplexes. 
We find that the resulting model is finite.}

\FullConference{3rd Quantum Gravity and Quantum Geometry School \\
		 February 28 - March 13, 2011\\
		 Zakopane, Poland}

\begin{document}

\section{Introduction}

Spin foam models can be viewed as discretised functional integrals for field theories of BF-type. These types of theories admit 
a formulation with a cosmological constant in three and  four space-time dimensions.
Models for theories with zero cosmological constant are based on the representation theory of simple Lie groups. 
Such models are given by infinite sums that diverge for a large class of manifolds. A natural regularisation is 
obtained by considering models based on the representation theory of quantum groups. In three and four space-time dimensions, 
the later models are believed to correspond to field theories with non-zero cosmological constant. 

A prototypical example of such a procedure is provided by the Turaev-Viro \cite{Turaev} regularisation of 
the Ponzano-Regge model \cite{ponzreg}.
The Ponzano-Regge model defines a functional integral for Euclidean three-dimensional gravity
with zero cosmological constant. It is given by an infinite weighted sum over all unitary, irreducible 
representations of the Lie 
group $\SU(2)$. Generically, this sum diverges and a natural regularisation is obtained using quantum groups. 
The idea is that $U_q(\su(2))$, the quantum deformation of $\SU(2)$, admits only a finite number of irreducible 
representations if the deformation parameter $q$ is chosen to be a root of unity.
This leads to a natural regularisation scheme for the Ponzano-Regge model; replace $\SU(2)$ in the definition of the 
model by its quantum deformation $U_q(\su(2))$ at root of unity. The resulting model defines a 3d TQFT called the 
Turaev-Viro model. Physically, the Turaev-Viro invariant is interpreted as a functional integral for Euclidean three-dimensional
gravity with positive cosmological constant $\Lambda$, if the deformation parameter $q$ is tuned to be a specific function of 
$\Lambda$. 

Following this intuition, it seems natural to follow the same procedure to regularise the potential divergences of 
higher dimensional models. This procedure has successfully been applied to the four-dimensional Ooguri \cite{ooguri} and 
Barrett-Crane \cite{BC,BC2} models in \cite{yetter2,yetter3} and \cite{yet4,PK} respectively. In \cite{Us}, we constructed 
an analysed a $q$-deformation of both Euclidean and Lorentzian versions of the EPRL model \cite{eprlmod}. Note also the 
independent work \cite{muxin}. In this paper, we 
will summarise the results obtained for the Lorentzian model. For further details or for results concerning the Euclidean 
model we refer the reader to the original paper \cite{Us}. 

\section{The quantum Lorentz group}

The model considered in this  paper is based on the representation theory of the quantum Lorentz group.

\subsection{Hopf algebra structures}

The quantum Lorentz group \cite{worpod,PE} is defined as the quantum double of $U_q(\su(2))$, where
$q = e^{-\kappa} \in \, ] 0,1[$ is a {\em real} deformation parameter.

\paragraph{The Hopf algebra $U_q(\su(2))$.}

We start by introducing the Hopf algebra $U_q(\su(2))$, adopting the conventions from \cite{PE}.
The Hopf algebra $U_q(\su(2))$ is the associative algebra generated multiplicatively by four generators
$q^{\pm J_z}$, $J_\pm$,  subject to the relations
\beq \label{rels1}
q^{\pm J_z}q^{\mp J_z}=1, \qquad q^{J_z}J_\pm
q^{-J_z}=q^{\pm1} J_\pm, \qquad
[J_+,J_-]=\frac{q^{2J_z}-q^{-2J_z}}{q-q^{-1}}.
\eeq
The comultiplication, counit and antipode are given by
\beqa
\label{uqcomult2} && \Delta(q^{\pm J_z}) =q^{\pm J_z} \otimes q^{\pm
J_z}, \;\;\;\;\;\;\;\; \Delta(J_\pm)=q^{-J_z} \otimes J_\pm+J_\pm \otimes q^{J_z},\\
\label{uqcounit2} && \epsilon(q^{\pm J_z})=1, \;\;\;\;\;\;\;\; \epsilon(J_\pm)=0,\\
\label{uqantip2} && S(q^{\pm J_z})=q^{\mp J_z}, \;\;\;\;\;\;\;\; S(J_\pm)=-q^{\pm
1} J_\pm,
\eeqa

The representation theory of  $U_q(\su(2))$
with a real deformation parameter $q$ closely resembles the representation theory of the Lie group $\SU(2)$. 
Irreducible finite-dimensional unitary representations are labeled by ``spins'' $I\in\mathbb{N}/2$.
As in the case of the Lie group $\SU(2)$,  the representation space $V_I$ of 
the irreducible representation $ \pi_{I}: U_q(\su(2)) \rightarrow\mathrm{End}(V_I)$  is $(2I+1)$-dimensional. 
The fusion rules for the tensor products $V_I \otimes V_J$ 
resemble the ones for the representations of  $\SU(2)$. We have
\beq
\label{fusionunity}
V_{I} \otimes V_{J} \cong  \bigoplus_{K = | I - J |}^{I + J} V_K,
\eeq
where the isomorphism $\cong$ is given by the Clebsch-Gordan intertwining operators
\beq
C^{K}_{\; IJ}: V_I \otimes V_J \rightarrow V_K, \hspace{3mm} \mbox{and} \hspace{3mm} 
C^{IJ}_{\;\, K} : V_K \rightarrow V_I \otimes V_J.
\eeq
As all multiplicities in \eqref{fusionunity} are equal to one, 
these intertwiners are unique up to normalisation. 
They  are non-zero if and only if  $I + J - K$, $J+K-I$ and $K+I-J$ are non-negative integers.
Their coefficients with respect to an orthonormal basis $\{e^I_m\}_{m=-I,...,I}$ of the complex vector space $V_I$ 
are the  Clebsch-Gordan coefficients
\beq
C^{IJ}_{\;\, K}(e^K_{m}) = \sum_{n,p} \left( \begin{array}{cc} n & p \\
                          I & J  \end{array} \right| \left. \begin{array}{c} K \\ m \end{array} \right)  
e^I_{n} \otimes e^J_{p}, \hspace{3mm} \mbox{and} \hspace{3mm}
C^{K}_{\; IJ}( e^I_{n} \otimes e^J_{p}) = \sum_{a} \left( \begin{array}{c} m \\ K
 \end{array} \right| \left. \begin{array}{cc} I & J \\ n & p \end{array} \right)  e^K_{m}.
\eeq

\paragraph{The Hopf algebra $F_q(\SU(2))$.}

The Hopf algebra $F_q(\SU(2))$ is the dual of the Hopf algebra $U_q(\su(2))$ and can be viewed as a quantum 
deformations of the algebra of polynomial functions on $\SU(2)$. 
A basis of $F_q(\SU(2))$ is given by the matrix elements $u^{\;\;m}_{I \;\; n}: U_q(\su(2))\rightarrow\C$ in the 
unitary irreducible representations of $U_q(\su(2))$
\beq
\label{matrixels}
u^{\;\;m}_{I \;\; n}(x)= e^{I m}\left(\pi_I(x)e^I_n\right), \;\;\;\; \forall x\in U_q(\su(2)).
\eeq
The duality pairing $\langle\,,\,\rangle: U_q(\su(2))\times F_q(\SU(2))\rightarrow \C$ is given by
\begin{align}\label{pair}
\langle x , u^{\;\;m}_{I \;\; n} \rangle = \pi_I(x)^{m}_{\;\; n}.
\end{align}
The Hopf algebra structure of $F_q(\SU(2))$ is induced by the one on $U_q(\su(2))$ via the pairing \eqref{pair}. 
In terms of the matrix elements $u^{\;\;m}_{I \;\; n}$, its algebra structure   is characterised by the relations 
\beq
\label{dualuqrels}
u^{\;\;m}_{I \;\; n} \,\cdot \, u^{\;\;p}_{J \;\; q} = \sum_{K,r,s} \left( \begin{array}{cc} m & p \\
                          I & J  \end{array} \right| \left. \begin{array}{c} K \\ r \end{array} \right)  
\, u^{\;\;r}_{K \;\; s} \, 
                          \left( \begin{array}{c} s \\ K
 \end{array} \right| \left. \begin{array}{cc} I & J \\ n & q \end{array} \right), \qquad\qquad 1=u^{\;\;0}_{0\;\;0}.
 \eeq                                                  
Its comultiplication, counit  and antipode take the form
\beqa
\label{coFq}
&& \Delta \left( u^{\;\;m}_{I \;\; n} \right) = \sum_p u^m_{I \;\; p} \otimes u^{\;\;p}_{I \;\; n},\\
&& \epsilon(u^{\;\;m}_{I \;\; n}) = \delta^{\;\;m}_{I \;\; n}, \label{counFq}\\
&& S(u^{\;\;m}_{I \;\; n}) = \epsilon_{I np} u^{\;\;p}_{I \;\; q} \epsilon_I^{-1qm},\label{antFq}
\eeqa
where $\delta^{\;\;m}_{I \;\; n}$ is the Kronecker symbol for the representation labeled by $I$ and the 
coefficients $\epsilon_{Inp}$ are the matrix elements of the bijective intertwiner $\epsilon_I : V_I \rightarrow V_I^*$ given 
explicitly by 
\beq\label{eps}
\epsilon_{I \, mn} = e^{-i \pi (I+m)} q^{I(I+1)} q^m\delta_{m,-n}.
\eeq

\paragraph{The quantum Lorentz group.}

The quantum Lorentz group is the quantum double of $U_q(\su(2))$
$$
D \hspace{0.3mm} (U_q(\su(2))) = U_q(\su(2)) \,  \hat{\otimes} \, F_q(\SU(2))^{op},
$$
where $F_q(\SU(2))^{op}$ is the Hopf algebra $F_q(\SU(2))$ with opposite coproduct, and the symbol
  `$\hat{\otimes}$'  indicates that the Hopf subalgebras $U_q(\su(2)) \otimes 1$ and $1 \otimes F_q(\SU(2))^{op}$ do 
not commute inside  $D \hspace{0.3mm} (U_q(\su(2)))$ but satisfy braided relations. See for instance \cite{PE}. As it is a 
quantum double, the quantum Lorentz group is a quasi-triangular Hopf algebra. We will 
describe the corresponding braiding later in the text.

The quantum double $D(U_q(\su(2)))$ is called the quantum Lorentz group because it is a quantum deformation of the universal 
enveloping algebra of the real Lie algebra $\mathfrak{sl}(2,\C)_{\mathbb{R}} \cong \mathfrak{so}(3,1)$. Indeed, the 
decomposition given above is the quantum analogue of the Iwasawa decomposition of the Lorentz algebra
$$
\mathfrak{sl}(2,\C)_{\mathbb{R}} \cong \su(2) \oplus \mathfrak{an}(2),
$$
where $\mathfrak{an}(2)$ is the Lie algebra of the Lie group
$$
\mathrm{AN}(2) = \left\{ \left( \begin{array}{cc} \lambda & n \\ 0 & \lambda^{-1} \end{array} \right) \;\; \mid 
\;\; \lambda \in \mathbb{R}^{*+}, \;\; n \in \mathbb{C} \right\}
$$
This is a direct consequence of the 
quantum duality principle which yields the identities 
$F_q(\SU(2)) = U_q(\mathfrak{su}(2))^* = U_q(\mathfrak{su}(2)^*) \cong U_q(\mathfrak{an}(2))$. Therefore, 
we will frequently use the notation 
\beq
D(U_q(\su(2)))= U_q(\mathfrak{sl}(2,\C)_{\mathbb{R}}).
\eeq

\subsection{Irreducible representations}

The irreducible unitary representations of $U_q(\mathfrak{sl}(2,\mathbb{C})_\mathbb{R})$  were first 
classified by Pusz \cite{Pusz}. In this paper, we will only consider  the representations of the principal series. These 
representations are labeled by a couple $\alpha=(n,p)$ with  $n\in \mathbb{Z}/2$ and $p\in [0,\frac{4 \pi}{\kappa} [$ or 
with $n=0$ and $p \in [0,\frac{2\pi}{\kappa} ]$.
We denote by $(\pi_{\alpha},V_{\alpha})$ the representation of $U_q(\mathfrak{sl}(2,\mathbb{C})_\mathbb{R})$ 
labeled by $\alpha$. It is a Harish-Chandra representation which decomposes into representations of 
$U_q(\mathfrak{su}(2))$ 
as follows 
\beq
\label{qrepresentation}
V_{\alpha} = \bigoplus_{I= \mid n \mid}^{\infty} V_I,
\eeq
where $V_I$ is the left $U_q(\mathfrak{su}(2))$-module introduced previously. A basis of the infinite dimensional 
vector space $V_{\alpha}$ is given by  $\{e^I_m \mid I\in \mathbb{N}, I\geq |n|, m = -I, ..., I \}$ where, 
for fixed $I$, $\{e^I_m\}_{m=-I,...,I}$ is the basis of $V_I$ introduced above. In terms of this 
basis, the action of $D(U_q(\su(2)))$ on the representation space $V_\alpha$ is given by 
the standard action of $U_q(\su(2))$ and the following action of $F_q(\su(2))$ 
\beq
\label{repQLGmod}
\pi_{\alpha}(u^{\;\;n}_{J \;\; n'}) \; e^L_p = \sum_{M,N}e^N_{p'} \; \left( \begin{array}{cc} p' & n \\
                          N & J  \end{array} \right| \left. \begin{array}{c} M \\ m \end{array} \right) 
                           \left( \begin{array}{c} m \\ M
\end{array} \right| \left. \begin{array}{cc} J & L \\ n' & p \end{array} \right) \; \Lambda^{JM}_{NL}(\alpha), 
\eeq
where $\Lambda^{JM}_{NL}(\alpha)$ are complex numbers defined in terms of analytic 
continuations of $6j$ symbols for $U_q(\su(2))$. As their expressions are lengthy and complicated, we will not 
give them here but refer the reader to \cite{PE}, where they are derived explicitly, and to \cite{PE2} where 
their properties are studied in details.

\section{The quantum EPRL intertwiner}

Given a triangulated $4$-manifold, there are three essential ingredients for the definition of a 4d spin foam model; 
the set of representations assigned to the triangles, the intertwining operators associated to the tetrahedra and the
amplitudes for the $4$-simplexes. In this section, we generalise the classical construction \cite{eprlmod} to the quantum 
group case.

\subsection{Quantum EPRL representations}

A quantum EPRL representation assigns a principal representation of the quantum Lorentz group to a representation
of the Hopf subalgebra $U_q(\su(2))$. This assignment depends on a fixed parameter $\gamma \in \R^+$, called the 
Immirzi parameter, and is defined as follows
\beq
K \mapsto (n(K),p(K)) := (K,\gamma K)
\eeq
Remark that for this assignment to produce principal representation, the representations of $U_q(\su(2))$ considered 
must be restricted to a specific subset of $\mathbb{N}/2$ since $p(K)$ must lie in $[0,\frac{4 \pi}{\kappa} [$.
This leads us to the following definition.

\begin{definition} {\bf (EPRL representations)}
Let $\mathcal{L} = \{ K \in \mathbb{N}/2 \, \mid \, 0 < K \leq 4 \pi / \gamma \kappa \}$ label a subset of representations of 
quantum $\SU(2)$. The Lorentzian EPRL representation  of  spin $K\in \mathcal{L}$ is the principal representation of the 
quantum Lorentz group labeled by
$
\alpha(K) = (n(K),p(K)) := (K,\gamma K). 
$
\end{definition}

Due to the restriction on the $U_q(\su(2))$ labels, the EPRL representation $\alpha(K) = (K, \gamma K)$ is a principal
representation of the quantum Lorentz group. It decomposes into quantum $\SU(2)$ representations as
\beq\label{decomp}
V_{\alpha(K)} \cong \bigoplus_{I=K}^{\infty} V_I.
\eeq 

\subsection{Quantum EPRL intertwiner}

The next step is to define the class of intertwining operators for the tetrahedra. Given a $n$-tuple 
$\alpha=(\alpha_1,...,\alpha_n)$ of principal representations of the classical Lorentz group $SL(2,\C)_{\R}$, a key 
ingredient appearing in the construction of the classical (i.e. non-deformed) EPRL intertwiner is the linear map
\beq\label{int}
\int_{SL(2,\C)_{\R}} dX \, \bigotimes_{i=1}^n \pi_{\alpha_i}(X) \;\;\;\; : \;\;\;\;\; V[\alpha]
\rightarrow V[\alpha],
\eeq
where $dX$ is a Haar measure on $SL(2,\C)_{\R}$ and $V[\alpha] = \bigotimes_{i=1}^n V_{\alpha_i}$.

This expression is generalised to the quantum group case by introducing a Haar measure (a biinvariant integral) on the
Hopf algebra $F_q(SL(2,\C)_{\R})$ dual to the quantum Lorentz group
$$
h : F_q(SL(2,\C)_{\R}) \rightarrow \C.
$$
The map $h$ is a linear form satisfying 
$$
(h \otimes id)\Delta(x) = h(x) 1, \;\;\;\; \mbox{and} \;\;\;\; (id \otimes h)\Delta(x) = h(x) 1,
\qquad \forall x \in F_q(SL(2,\C)_{\R}).
$$
These identities imply that $h$ is invariant under the left- and right-action of the quantum Lorentz group on 
its dual Hopf algebra $F_q(SL(2,\C)_{\R})$.

Let $\{ x_A \}_{A}$ be a basis of $F_q(SL(2,\C)_{\R})$ and introduce the dual basis $\{ x^A \}_A$ of the quantum Lorentz
group. Expression \eqref{int} is generalised as follows
\beq\label{T}
T_{[\alpha]} = \sum_A \bigotimes_{i=1}^n \pi_{\alpha_i}(\Delta^{(n-1)}(x^A)) h(x_A),
\eeq
where $\Delta^{(n)}$ denotes the $n$-fold coproduct in the quantum Lorentz group
$$\Delta^{(n)}=(\Delta\otimes \text{id}^{\otimes (n-1)}) \circ \Delta^{(n-1)} \;\;\;\; \mbox{for} 
\;\; n>1, \;\;\;\; \Delta^{(1)}=\Delta.$$

Now, consider an EPRL representation $\alpha(K)$ and the projection map $f_{\alpha}^K : V_{\alpha} \rightarrow V_K$ 
associated to the lowest weight factor in the decomposition \eqref{decomp}.
The dual of this map induces an embedding
\beq
f^* : \Hom_{U_q(\su(2))}\left(\bigotimes_{i=1}^n V_{K_i} , \C \right) \rightarrow \Hom_{U_q(\s(2,\C)_{\R})}\left(
\bigotimes_{i=1}^n V_{\alpha_i(K_i)},\C \right).
\eeq

Using the above, we define a quantum EPRL intertwiner as follows. 

\begin{definition} {\bf (Quantum EPRL intertwiner)}
Let $K = (K_1, ..., K_n)$ be a $n$-tuple of elements of $\mathcal{L}$ and $V[K] = \bigotimes_{i=1}^n V_{K_i}$ 
be 
the corresponding representation space of $U_q(\su(2))$. Denote by  $\alpha = (\alpha_1(K_1),...,\alpha_n(K_n))$ the 
associated $n$-tuple  of EPRL representations and by 
$V[\alpha] = \bigotimes_{i=1}^n V_{\alpha_i}$ the tensor product of their  representation spaces. The quantum EPRL 
intertwiner $\iota_{[\alpha]} = f^*(\Lambda_{[K]})$ associated to an intertwiner $\Lambda_{[K]}$  in  
$\mathrm{Hom}_{U_q(\su(2))}(V[K], \C)$ is  the linear  map $\iota_{[\alpha]} : V[\alpha] \rightarrow \C$ defined by  
\beq
\label{quintertwiner}
\iota_{[\alpha]}= \sum_{A}  \Lambda_{[K]} \circ \bigotimes_{i=1}^n f^{K_i}_{\alpha_i} \circ 
\left( \bigotimes_{i=1}^n \pi_{\alpha_i(K_i)} 
(\Delta^{(n-1)}(x^A)) \right) h(x_{A}).
\eeq
\end{definition}

As this definition involves an infinite sum, it has to be established that the quantum EPRL
intertwiner is well-defined. In the following we  will mainly be interested in the case $n=4$. In this case, an orthogonal
basis of the vector space $\Hom_{U_q(\su(2))}\left(\bigotimes_{i=1}^4 V_{K_i} , \C \right)$ is given by:
$$
\{ \lambda_{[K],I}\}_{I \in \mathbb{N}/2}, \;\;\;\;\;\; \lambda_{[K],I} = d_I \circ (C_{K_1 K_2 I} \otimes C^I_{\;K_3 K_4}).
$$
Here, $C_{IJK} = \epsilon_K \circ C^K_{\;IJ}$ with $\epsilon_K$ defined in \eqref{eps}, and 
$d_I : V_I^* \otimes V_I \rightarrow
\C$; $e^{Ia} \otimes e^I_b \mapsto \delta^a_b$, is the evaluation map in the representation category of quantum $\SU(2)$. 
We now state an important convergence result.

\begin{theorem}
Let  $\lambda_{[K],N}$ be an element of the basis $\{\lambda_{[K],N}\}_{N}$
and  $e^L_{c}[\alpha] = \bigotimes_{i=1}^4 e^{L_i}_{c_i}(\alpha_i)$ a basis of $V[\alpha]$. The 
evaluation of the quantum EPRL intertwiner $\iota_{[\alpha],N} = f^*(\lambda_{[K],N})$ is given by
\beqa
\label{evalL}
&& \iota_{[\alpha],N} \, (e^L_{c}[\alpha]) 
= \sum_{I} \sum_{M_1,...,M_4} \!\!\![2I+1]_q \, q^{-2 b_4} \delta^{b'_1}_{\;\; b_4}
\Lambda^{I M_1}_{K_1L_1}(\alpha_1) 
\Lambda^{I M_2}_{K_2L_2}(\alpha_2)  \Lambda^{I M_3}_{K_3L_3}(\alpha_3)  \Lambda^{I M_4}_{K_4L_4}(\alpha_4)
\nn \\
&& \left( \begin{array}{cc} a_1 & b_1 \\ K_1 & I  \end{array} \right| \left. \begin{array}{c} M_1 \\ d_1 
\end{array} \right) \left( \begin{array}{c} d_1 \\ M_1 \end{array} \right| \left. 
\begin{array}{cc} I & L_1 \\ b'_1 & c_1 \end{array} \right) \left( \begin{array}{cc} a_2 & b_2 \\
K_2 & I  \end{array} \right| \left. \begin{array}{c} M_2 \\ d_2 \end{array} \right)\!  
\left( \begin{array}{c} d_2 \\ M_2
 \end{array} \right| \left. \begin{array}{cc} I & L_2 \\ b_1 & c_1 \end{array} \right)\nn \\
&& \left( \begin{array}{cc} a_3 & b_3 \\ K_3 & I  \end{array} \right| \left. \begin{array}{c} m_3 \\ d_3 
\end{array} \right) \left( \begin{array}{c} d_3 \\ m_3 \end{array} \right| \left. 
\begin{array}{cc} I & L_3 \\ b_2 & c_3 \end{array} \right) \left( \begin{array}{cc} a_4 & b_4 \\
K_4 & I  \end{array} \right| \left. \begin{array}{c} M_4 \\ d_4 \end{array} \right)                          
\left( \begin{array}{c} d_4 \\ M_4 \end{array} \right| \left. \begin{array}{cc} I & L_4 \\ b_3 & c_4 \end{array} \right) 
\nn \\
&& \lambda_{[K],N}(e^{K_1}_{a_1} \otimes e^{K_2}_{a_2} \otimes e^{K_3}_{a_3} \otimes e^{K_4}_{a_4}), \nn
\eeqa
where $[n]_q = (q^n - q^{-n}) / (q - q^{-1})$ denotes a $q$-number.
This multiple series converges absolutely.
\end{theorem}

This theorem implies that the defined $q$-EPRL intertwiner is well-defined.
Another important property of the $q$-EPRL intertwiner is its behaviour under braiding. The representation category of the 
quantum Lorentz group is a braided tensor category (with infinite dimensional objects) because the quantum Lorentz group is 
a quasi-triangular Hopf algebra. This means that there exists an $R$-matrix $R \in D(U_q(\su(2))) \otimes D(U_q(\su(2)))$ 
which is an immediate consequence of the fact that we are working with a quantum double. From this $R$-matrix, one can 
construct an intertwining operator 
$c_{\alpha_1, \alpha_2}: V_{\alpha_1} \otimes V_{\alpha_2} \rightarrow V_{\alpha_2} \otimes V_{\alpha_1}$
called a braiding. This operator is given by
\beq
c_{\alpha_1, \alpha_2} = \tau_{\alpha_1,\alpha_2} \circ \pi_{\alpha_1} \otimes \pi_{\alpha_2}(R),
\eeq
where $\tau : x \otimes y \mapsto y \otimes x$ is the flip map. Using the explicit form of the $R$-matrix \cite{PE2} and 
the action of the quantum Lorentz group on the module $V_{\alpha}$, it is immediate to obtain the following expression for 
the action of the braiding
\beq
\label{repR}
c_{\alpha,\beta} \, (e^I_c \otimes e^J_d) = \sum_{K} 
\sum_{L} e^L_f \otimes e^I_e \; 
 \left( \begin{array}{cc} f & e \\
                          L & I  \end{array} \right| \left. \begin{array}{c} K \\ g \end{array} \right)          
                  \left( \begin{array}{c} g \\ K
 \end{array} \right| \left. \begin{array}{cc} I & J \\ c & d \end{array} \right) \; \Lambda^{IK}_{LJ}(\alpha).
 \eeq
Note that although these sums are infinite, there is only a finite number of 
non-zero terms \cite{PE2}. Consequently, there are no  issues with convergence. 
We are now ready to state the following result.

\begin{proposition}
The $q$-EPRL intertwiner $\iota_{[\alpha],J} = f^*(\lambda_{[K],J})$ transforms as
$$
\iota_{\alpha_1(K_1)\alpha_2(K_2)\alpha_3(K_3)\alpha_4(K_4),J} \circ c_{\alpha_2,\alpha_1} = \sum_K 
\Lambda^{K_2J}_{K_1K}(\alpha_2)
\; \iota_{\alpha_2(K_2) \; \alpha_1(K) \alpha_3(K_3)\alpha_4(K_4),J},
$$
and is thus not invariant under braiding.
\end{proposition}

Note\footnote{Note also the abuse of notation in the right hand side of the above equation due to the fact that 
$\alpha_1(K)$ is
not necessarily an EPRL representation. The notation is nevertheless used for notational compacity.} that this is sharp 
contrast
with the case of the quantum deformation of the Barrett-Crane intertwiner \cite{yet4,PK} 
which is invariant under braiding.

\section{The $4$-simplex amplitude}

We are now ready to construct the amplitude for $4$-simplexes labeled by EPRL representations and $q$-EPRL intertwiners. 
Such an amplitude is defined with the aid of the graphical calculus of spin networks.
There are two main difficulties in the definition of the amplitude that arise from the fact that the representation spaces 
of the EPRL representations are infinite-dimensional. The first is that there is no coevaluation map that intertwines 
the trivial representation of the quantum Lorentz group on $\C$ with a representation on the tensor product 
$V_\alpha\otimes V_\alpha^*$ and therefore no notion of a quantum trace. The second difficulty is that a naive definition 
of the amplitude for the four-simplexes gives an infinite answer.

These problems  arise in a similar fashion in the classical Lorentzian BC \cite{BC2} and EPRL \cite{eprlmod} models. A
solution to the first problem was provided in \cite{notts3} where a Lorentzian graphical calculus based on non-invariant
tensors and bilinear forms was invented. A regularisation prescription that circumvents the second difficulty has been 
given in
\cite{BB} and \cite{roberto} for the BC and EPRL models, respectively. 
Extending these procedures to the quantum Lorentz group, we will overcome these issues and consistently construct a finite 
amplitude for the $4$-simplexes.

\subsection{Graphical calculus}

The graphical calculus is defined by associating certain algebraic quantities to diagrams drawn in the plane. We first 
describe the algebraic side of the calculus before relating the algebra to the diagrams.

\subsubsection{EPRL tensors and invariant bilinear form}

We first need a notion of dual quantum EPRL intertwiners. These dual objects are required since we cannot pair 
$q$-EPRL intertwiners together because of the absence of a coevaluation map.
As in the Lie group case, the vector space  
$\bigotimes_{i=1}^n V_{\alpha_i}$ does not contain 
tensors that are invariant under the action of the quantum Lorentz group and such objects do not exist per se. Accordingly, 
dual $q$-EPRL intertwiners  are replaced by non-invariant quantities which can be viewed as the quantum group analogue of  
the tensors considered in \cite{notts3}. These quantities, which are referred to as vertex functions in \cite{PK},  will be 
called EPRL tensors in the following.

\begin{definition} {\bf (Quantum EPRL tensor)}
Let $K = (K_1, ..., K_n)$ be a $n$-tuple of representations of $U_q(\su(2))$ labeled by elements of $\mathcal{L}$. Denote 
by $\alpha= (\alpha_1(K_1),...,\alpha_n(K_{n}))$ the associated $n$-tuple of EPRL representations, 
and consider an element  $\Lambda^{[K]} \in \mathrm{Hom}_{U_q(\su(2))}(\C, V[K])$.  The quantum EPRL tensor 
$\Psi^{[\alpha]}$ associated to $\Lambda^{[K]}$ is  defined by 
\beq
\label{tensor}
\Psi^{[\alpha]} = \left[ \sum_{A}  \left( \bigotimes_{i=1}^n \pi_{{\alpha}_{i(K_i)}} \left( \Delta^{(n-1)}(x^A) \right)
\right) \circ \bigotimes_{i=1}^n f_{K_i}^{\alpha_i} \circ \Lambda^{[K]} \right] \otimes x_{A},
\eeq
where $f_K^{\alpha} : V_K \rightarrow V_{\alpha}$ is the inclusion map associated to the direct sum \eqref{decomp}.
The vector space of EPRL tensors associated with $[\alpha]$ is the vector space  $H[\alpha] := V[\alpha]\otimes 
F_q(\SL(2,\C)_{\mathbb{R}})$.
\end{definition}

The second algebraic object required for the calculus is an invariant bilinear form with which one can pair $q$-EPRL tensors.

\begin{lemma} 
\label{lemma}
Let $V_{\alpha}^* = \bigoplus_{I=m}^\infty V_I^*$ be the dual to the vector space $V_{\alpha}$.
There exist a bijective intertwiner $\phi^{\alpha} : V_{\alpha} \rightarrow V_{\alpha}^*$ whose expression with respect to 
the 
basis 
$\{e^I_a\}_{I,a}$ of $V_{\alpha}$ and the dual basis $\{e^{Ia}\}_{I,a}$ of $V_{\alpha}^{*}$ is given by
\beq 
\label{Lorint}
\phi^\alpha (e^I_a)=  \phi^{\alpha}_{IaJb} e^{Jb} \qquad \phi^{\alpha}_{IaJ_b} = 
c_{\alpha} q^{-I(I+1)} \delta_{IJ} \epsilon_{I ba},
\eeq
where  $c_{\alpha}$ is a constant and $\epsilon_{Iab}$ is given by \eqref{eps}.
The bilinear form 
$\beta_{\alpha} : V_{\alpha} \otimes V_{\alpha} \rightarrow \C$ 
\beq
\label{bilinear}
 \beta_{\alpha}(v,w) = \phi^{\alpha}(w)(v), \qquad \forall v,w \in V_{\alpha}, 
\eeq
satisfies the invariance property $\beta_{\alpha}( v , \pi_{\alpha}(a) w) = \beta_{\alpha}( \pi_{\alpha}(S(a)) v ,  w)$ 
for all $a\in D(U_q(\su(2)))$.
\end{lemma}


\subsubsection{Graphical calculus}

\begin{figure}
  \includegraphics[scale=0.35]{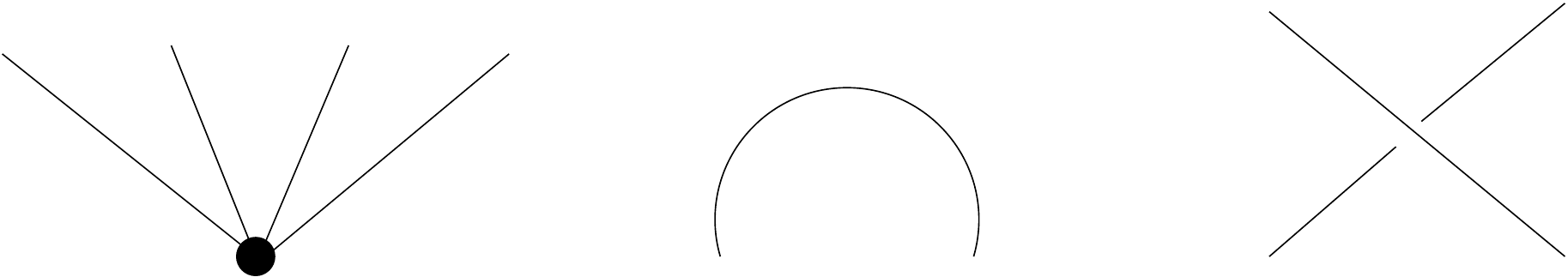}
\centering  \caption{Elements of the graphical calculus: $q$-EPRL tensor, bilinear form and braiding.}
  \label{calculus}
\end{figure}

The elements of the graphical calculus are vertices, arcs and crossings.
The diagram corresponding to a $q$-EPRL tensor $\Psi^{[\alpha]}$ is a vertex, to which is attached a basis element of 
$F_q(\SL(2,\C)_{\R})$, with $n$ lines coloured by elements
of the $n$-tuple $[\alpha]$. Arrows do not appear on the diagrams and the convention is that
all lines are pointing away from the vertex. The tensor product of $p$ EPRL tensors 
$\Psi^{[\alpha_1]} \otimes ... \otimes \Psi^{[\alpha_p]}$ is given 
by drawing the $p$ vertices on a horizontal line, the order in the tensor product being read from left to right. 

The second ingredient for the graphical calculus enables us to pair EPRL tensors. The pairing is defined in terms of  
the invariant bilinear form $\beta_{\alpha} : V_{\alpha} \otimes V_{\alpha} \rightarrow \C$ given above. In the diagrams, 
this bilinear form is depicted by an arc\footnote{Note that in contrast to the diagrams for ribbon categories, the upward 
and downward arcs have no direct meaning. The arcs are free to go up and down between the vertices.} with which one can 
connect two different lines coloured by the same representation.

When more than one pair of edges are paired in a diagram, crossings  can occur. To each crossing 
where the left-hand leg goes under the right-hand leg, we associate the third ingredient of the graphical calculus; 
a braiding $c_{\alpha_2, \alpha_1}$. 

An important class of diagrams are closed diagrams.  
A closed diagram $\Gamma$ with $p$ vertices corresponds to an element $\phi(\Gamma)$ of 
$\End(\C) \otimes F_q(\SL(2,\C))^{\otimes p} \cong F_q(\SL(2,\C))^{\otimes p}$.
The evaluation $ev(\Gamma)$ of a closed diagram $\Gamma$ is then defined via the Haar integral in the spirit of Feynman 
diagram evaluations. 

The naive evaluation of a closed diagram with $p$ vertices would correspond to setting  
$ev(\Gamma)=h^{\otimes p} (\phi(\Gamma))$. However, such an evaluation is generically divergent and 
needs to be regularised.  
This is done in analogy to the classical case \cite{BB,roberto} by removing the Haar measure or integration at one 
(randomly chosen) vertex as in \cite{PK}. The invariance of the Haar integral implies that the result is independent of 
the chosen vertex. Moreover, it  implies  that $(h^{\otimes p - 1} \otimes id) (\phi(\Gamma)) = ev(\Gamma) 1$, where 
$ev(\Gamma)$ is a complex number or infinity and 1 is the unit in $F_q(SL(2,\C))$.  
The evaluation of $\Gamma$ is therefore obtained by applying $p-1$ copies of the Haar 
measure to $\Phi(\Gamma)$ and then applying the counit of $F_q(SL(2,\C))$  to the resulting expression
$$
ev(\Gamma) = \epsilon \left( (h^{\otimes p - 1} \otimes id) (\phi(\Gamma)) \right).
$$
If the result is finite, the diagram $\Gamma$ is said to be integrable. 

\subsection{Amplitude for the $4$-simplexes}

We are now ready to define the amplitude for the four-simplexes.
Let $M$ be an oriented, closed  triangulated $4$-manifold. We will note $\Delta_{(2)}$, $\Delta_{(3)}$ and $\Delta_{(4)}$ 
the sets of triangles, tetrahedra and $4$-simplexes of $M$ respectively. 

A colouring of $M$ is a map $\alpha : \Delta_{(2)} \rightarrow \mathrm{Irrep} \; U_q(\s(2,\C)_{\R})$; 
$\Delta \mapsto \alpha_{\Delta}$, that associates an EPRL representation to each oriented triangle $\Delta$ of $M$.
Given a coloured triangulated manifold $M$, we define, for every oriented tetrahedron $t \in \Delta_{(3)}$, the state space
$$
H_t = \left( \bigotimes_{\Delta \in \partial t} V_{\alpha_{\Delta}} \right) \otimes F_q(\SL(2,\C)_{\R}).
$$
A state is an assignment of a $q$-EPRL tensor $\Psi_t \in H_t$ to each tetrahedron $t$ of $M$.

\begin{figure}
  \includegraphics[scale=0.25]{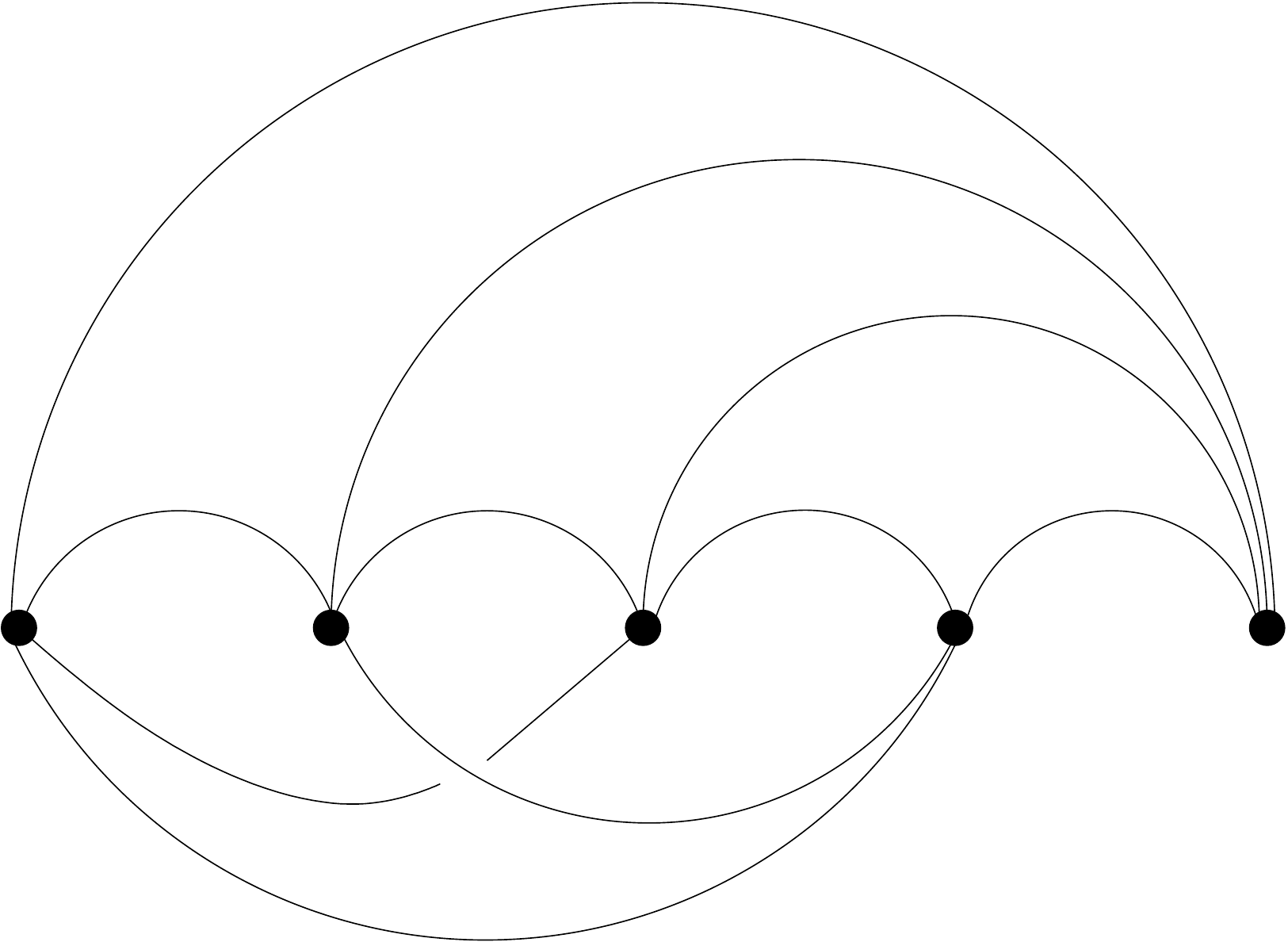}
\centering  \caption{Diagram $\Gamma_5$}
  \label{amplitude}
\end{figure}

The amplitude for a $4$-simplex $\sigma$ in $\Delta_{(4)}$ is then a linear map
\beq
A_{\sigma} : \bigotimes_{t \in \partial \sigma} H_t \rightarrow \C; \;\;\;\;\;\; 
A_{\sigma}(\Psi_1 \otimes ... \otimes \Psi_5) = ev(\Gamma_5),
\eeq
where the diagram $\Gamma_5$ is depicted in figure \ref{amplitude} and we have labeled the tetrahedra of $\partial \sigma$ 
with labels from one to five.

In \cite{Us}, we proved the following theorem which ensures that the evaluation of the $\Gamma_5$ diagram is well-defined, 
i.~e.~that the four-simplex amplitude is finite. 

\begin{theorem} The four-simplex amplitude $ev(\Gamma_5)$ converges absolutely.
\label{ampconv}
\end{theorem}

\section{The quantum spin foam model}

Using the notations and definitions from the previous section, we can now define the partition function for the quantum 
EPRL model associated to a closed, oriented triangulated manifold $M$:
\beq
\mathcal{Z}(M,\gamma,q) = \sum_{K,J} \prod_{\Delta} [2K_{\Delta} + 1]_q \prod_{\sigma} 
A_{\sigma} \left(\alpha_{\Delta},\Psi_{\sigma}(J) \right).
\eeq
Here, $q \in \mathbb{R}$ is the deformation parameter, the sum ranges over all $K$ in $\mathcal{L}$ and over the elements of a 
basis of $U_q(\su(2))$-intertwiners $(\lambda_{[K],J})_J$ entering the definition of the EPRL tensors for each tetrahedron $t$ 
of $M$. 
The state $\Psi_{\sigma}$ for the $4$-simplex $\sigma$ is given by 
$\Psi_{\sigma} = \bigotimes_{t \in \partial \sigma} \Psi_{t}$, where $\Psi_{t}$ 
is the state associated to the tetrahedron $t$. The products run over all the triangles $\Delta$ and $4$-simplexes 
$\sigma$ of $M$.

The weight associated to the triangles is fixed from gluing arguments as in the classical case \cite{Carlo2}. As there 
is only a finite number of representations in the label set  $\mathcal{L}$, the sum involves only a 
finite number of terms and hence converges. Given the convergence of the EPRL intertwiners and the  $4$-simplex amplitude 
for fixed labels $K$, the convergence of the Lorentzian $q$-EPRL partition function is therefore immediate for all closed 
triangulated manifolds $M$.

\section{Discussion}

We conclude with a remark on the physical interpretation of the quantum deformation presented in this paper.
The model that we have considered is a $q$-deformed version of the EPRL spin foam model. As the latter is a model for 
quantum gravity with vanishing cosmological constant, it is plausible that  the $q$-deformed model should
describe some aspects of four-dimensional quantum gravity in Lorentzian de-Sitter space. Its deformation 
parameter should then be related to a positive cosmological constant $\Lambda$ via
 $$
q = \exp (- l_p^2 / l_c^2),
$$
where $l_p$ is the Planck length and $l_c=1/\sqrt{\Lambda}$ the cosmological length. 

Interestingly, this relation leads to a bound on the area spectrum. 
The area spectrum  for a triangle $\Delta$ of $M$ that is coloured by a representation $\alpha(K)$ is given by
$$
A(\Delta) = 8 \pi l_p^2 \gamma \sqrt{K(K+1)}.
$$
With the relation between  the deformation parameter $q$ and the cosmological constant given above, one obtains a bound on 
this spectrum in terms of the cosmological length $l_c$ in the regime where $l_P<<l_c$:
\beq
A(\Delta) \leq 32 \pi^2 l_c^2.
\eeq

\subsection*{Aknowledgments}

We would like to thank the organisers of 3rd Quantum Gravity and Quantum Geometry School (Zakopane, February 28 - March 13, 
2011) where these results were presented.
Both authors acknowledge support from the Emmy Noether grant ME 3425/1-1 of the German Research foundation (DFG).

\end{document}